\documentclass[conference,letterpaper]{IEEEtran}
\usepackage[utf8]{inputenc}
\usepackage{amsthm}
\usepackage[T1]{fontenc}
\usepackage{url}
\usepackage{cite}
\usepackage[cmex10]{amsmath}
\newtheorem{theorem}{Theorem}[section]
\newtheorem{corollary}{Corollary}[theorem]
\newtheorem{lemma}[theorem]{Lemma}
\theoremstyle{definition}

\usepackage[ruled,vlined]{algorithm2e}

\usepackage{algpseudocode}%
\usepackage{color}
\usepackage{amssymb}
\usepackage{subfig}
\usepackage{graphicx}
\usepackage{wrapfig}

\date{\today}

\newcommand{\E}{{\bf{\mathsf{E}}}}
\newcommand{\KL}{\mathrm{KL}}

\newcommand{\NORMAL}{\mathsf{Normal}}

\newcommand{\LAP}{\mathsf{Lap}}

\begin{document}
\title{Hiding Signal Strength Interference from Outside Adversaries}
\author{%
  \IEEEauthorblockN{Mingxuan Han}
  \IEEEauthorblockA{University of Utah\\ School of Computing\\
                    Email: u1209601@umail.utah.edu}
  \and
  \IEEEauthorblockN{Jeff M. Phillips}
  \IEEEauthorblockA{University of Utah\\ School of Computing\\
                    Email: jeffp@cs.utah.edu}
  \and 
  \IEEEauthorblockN{Sneha Kumar Kasera}
  \IEEEauthorblockA{University of Utah\\ School of Computing\\  Email: kasera@cs.utah.edu}
}
\maketitle
\begin{abstract}
    The presence of people can be detected by passively observing the signal strength of Wifi and related forms of communication.  This paper tackles the question of how and when can this be prevented by adjustments to the transmitted signal strength, and other similar measures.  The main contribution of this paper is a formal framework to analyze this problem, and the identification of several scenarios and corresponding protocols which can prevent or limit the inference from passive signal strength snooping.  
\end{abstract}


\section{Introduction}

Wifi is the dominant last leg of connecting devices to the internet, and is ubiquitous in households and businesses.  However, recent work~\cite{moving_average,RFsensor,LOCAL_FREE} has shown that these wifi signals can leak the location of occupants in those homes and businesses, even if the occupants are passive and do not have devices sending signals.  The presence of a person can reduce the Wifi signal strength, and an adversary outside of a home or business can detect this reduction, inferring the presence and even location of the person.  

This paper examines if and when it is possible to prevent the leaking of the presence of a person using only passive reduced signal strength of Wifi and other wireless transmissions. Different from, for instance,~\cite{Bloch_et_al,Bash_et_al}, we assume the messages can be encrypted, and for high security settings equipment can even continually fill the channel with messages in a regular (or random) pattern to avoid detection of the presence of signals. The only information leaked is the magnitude of the signals. 


This is a challenging domain with many possible strategies an adversary could use~\cite{SRD,BBDRKB,zhu2018wireless,monitor_breathe}.  
As such, this paper starts with a very simple model where a formal analysis can be developed.  Then it builds on these basic ideas to create a more comprehensive array of possible modeling assumptions.  The first model is in 1 dimension where the sender has full information, then when the model does not know if a person interferes or not.  We ultimately consider 2-dimensional (spatial) models, and if the sender knows the state of potentially interfering people or adversaries.  We empirically demonstrate the effectiveness of our models in the simple controlled scenarios.  
While these models do not reach the specification of the transmission protocol and hardware devices, they develop a characterization of which factors are essential to models, and which are less pertinent.  

Hence, the main contribution of this paper is \emph{the formalization of how to protect against signal strength inference attacks}, which outcomes are possible, and a series of modeling assumptions and corresponding protocols and analysis to protect against inference. Our methods are based on information theoretic and statistical information arguments, and given modeling assumptions are impervious to any adversarial attack, or show  a certain amount of transmissions must be made before the presensence of the person can be identified with sufficient (e.g., $95\%$) confidence.

\subsection{Related Works}
Previous related work mostly provides methods to detect the presence of people using RSS or similar signals.  This includes work based on the moving average~\cite{moving_average}, and similar to our noisy model, claims there is detection if the moving average is greater than some threshold.  
Abid \emph{et.al.}~\cite{AKKM} shows that witrack, a system
that tracks the 3D motion of a user from the radio signals reflected off a body, can localize the center of a human body to within a about 10 to 13 cm. 
This group also shows~\cite{FD} that they can track a human by treating the motion of a human body as an antenna array and tracking the resulting RF beam and show how one can use MIMO (Multiple Input Multiple Output) interference nulling to eliminate reflections off static objects and focus the receiver on a moving target. 

Another line of work related to statistical inference in signal strength~\cite{RFBO} estimates the parameters of a single-frequency complex tone from a finite number of noisy discrete-time observations.  Moschitta \emph{et.al.{}}~\cite{CRLBPEQS} provide a Cramer-Rao Lower Bound (CRLB) for the parametric estimation of quantized sinewaves. Similarly, Abrar \emph{et.al.{}}~\cite{abrar2020quantifying} contributes a CRLB on an attacker’s monitoring performance as a function of the RSS step size and sampling frequency.  
Similar models use change point detection statistics~\cite{TOV}, information theory and signal processing problem~\cite{EIT, FSSP}, or machine learning~\cite{random_forest,deep_learning}. 

In contrast our work provides a rigorous framework for characterizing when a person might or statistically \emph{cannot} be detected by \emph{any} method, or bounds the rate of potential detection up to some statistical confidence.  
A different tact to characterizing when detection is or is not possible uses game-theoretic approaches~\cite{game_theory_1, game_theory_2, game_theory_3}, as opposed to our statistical/information theory approach.


\section{Structural Properties}
\label{sec:structural}

We begin with some basic properties about distributions which will guide our characterization of various scenarios and the corresponding protocols.  The ``signal'' is a bit $b$, if a person interferes with a signal ($b=1$) or not ($b=0$).  
We identify three cases: when the scenarios are completely indistinguishable, when it must be one of the scenarios and not the other, and when it is not immediately clear, but the adversary gains information favoring one or the other.  In this last scenario, we quantify how much information the adversary gains, and then if $n$ readings are made in an iid fashion, when the adversary can reach a certain amount of confidence about one scenario or the other.  This reduces to the expectation under a distribution $\mu$, denoted $\E_\mu$, of the  Kullback-Liebler (KL) divergence between certain distributions, denoted $\KL(\cdot \mid \mid \cdot)$. 

\begin{lemma} \label{lemma-struct}
Consider a set $X$ of observations from one of the two distributions $f_1(\cdot)$ amd $f_2(\cdot)$, which are characterized by different parameters say $\mu_1$, $\mu_2$. If we are interested in either $X \sim f_1(x \mid \mu_1)$ or $X \sim f_2(x \mid \mu_2)$, the problem can be divided into three cases:

\begin{itemize}
    \item Case 1 (perfect hiding):  If $\mu_1 = \mu_2$ then naturally $X \sim f(X \mid \mu_1) \buildrel d \over = f(X \mid \mu_2)$ and we cannot distinguish that $X$ are drawn from $f(X \mid \mu_1)$ or $f(X \mid \mu_2)$.
    \item Case 2 (noisy hiding): If $\mu_1 \ne \mu_2$ and the logarithm of the likelihood ratio satisfies 
    \begin{align*}
       M = \ln (\frac{L(X \mid \mu_1)}{L(X \mid \mu_2)}) \le \ln(\frac{1-p}{p}),
    \end{align*}
   we can distinguish with more than $1-p$ confidence that $X$ is from $f_1(X \mid \mu_1)$ rather than $f_2(X \mid \mu_2)$.  If $X \sim \mu_1$, then this holds in expectation if it contains $n$ iid samples, and $\E_{\mu_1}(M) = n \cdot \KL(f_1(\mu_1) \mid \mid f_2(\mu_2))$, only if $n > \frac{\ln \frac{1-p}{p}}{\KL(f_1(X \mid \mu_1) \mid \mid f_2(X \mid \mu_2))}$.
    \item Case 3 (immediate detection): If for any observation $x \in X$, $\text{L}(x \mid \mu_2) = 0$ or $\text{L}(x \mid \mu_1) = 0$, we can immediately distinguish from which distribution $X$ is drawn. 
\end{itemize}
\end{lemma}

\begin{proof}
Case 1 and 3 are immediate.  
%
For the Case 2 the logarithm of the likelihood ratio is defined as 
$
    M = \ln(\frac{L(X \mid \mu_1)}{L(X \mid \mu_2)}).
$
If the $X \sim \mu_1$ then
\begin{align*}
    \E_{\mu_1}(M) = \E_{\mu_1}\left(\sum_{i = 1}^{n} \ln(\frac{L(x_i \mid \mu_1)}{L(x_i \mid \mu_2)})\right).
\end{align*}
Since $x_i$ are i.i.d. and by the definition of KL-divergence, 
\begin{align*}
    \E_{\mu_1}(M) &= n \cdot \E_{\mu_1}(\ln(\frac{L(x_i \mid  \mu_1)}{L(x_i \mid \mu_2)}), \\
                &= n \cdot \KL(f_1(x \mid \mu_1) \mid \mid f_2(x \mid \mu_2)).
\end{align*}
We can understand the likelihood ratio by normalizing the numerator, denominator pair so $\ln \frac{1-p}{p}$; then if $p = 0.05$ then we can think the confidence that $X$ comes from $f_1(x \mid \mu_1)$ is $0.95$ and the confidence that $X$ comes from $f_2(x \mid \mu_2)$ is $0.05$.  Hence, if $M \le \ln((1-p)/p)$ then with at most $1 - p$ confidence that $X$ are drawn from $f_1(x \mid \mu_1)$. Hence by contrapositive, in expectation only if $n > \frac{\ln \frac{1-p}{p}}{\KL(f_1(x \mid \mu_1) \mid \mid f_2(x \mid \mu_2))}$ can we say with more than $1 - p$ confidence that $X \sim f_1(x \mid \mu_1)$.
%
\end{proof}


\subsection{Special cases on the Case 2: Noisy Hiding}
Now we analyze two specifications of Case 2 in Lemma \ref{lemma-struct}: when $f_1$ and $f_2$ are Laplace or Normal.  
Recall $\LAP(\mu,\sigma)$ at value $x$ has the form $\exp(-|x-\mu|/\sigma)$, and $\NORMAL(\mu,\sigma)$ at value $x$ has the form $\exp(-(x-\mu)^2/2\sigma^2)$, each with a different normalizing constant, which will not play a role in our analysis, so is omitted for simplicity.

\begin{corollary}\label{laplace-corollary}
Consider when a set $X$ of observations is from either $\LAP(\mu_1, \sigma)$ or $\LAP(\mu_2, \sigma)$, where $\mu_2 - \mu_1 = \delta > 0$.   If $X$ is from $\LAP(\mu_1, \sigma)$, then at least $n > \ln(\frac{1 - p}{p})\frac{\sigma}{\delta}$ iid observations are needed for 
than $(1-p) \cdot 100\%$ confidence.  
\end{corollary}
\begin{proof}
According to Case 2 in Lemma \ref{lemma-struct}, we have 
\begin{align*}
    M &= \ln(\frac{L(X \mid \mu_1)}{L(X \mid \mu_2)}) 
      = \sum_{i = 1}^{n} \ln  \frac{ \exp( - \frac{|x_{i} - \mu_{1}|}{\sigma})}{\exp( - \frac{|x_{i} - \mu_{2}|}{\sigma})}. 
\\ &= 
\sum_{i=1}^n  \ln \left(\exp(-|x_i-\mu_1|/\sigma) \right) + \ln \left(\exp(|x_i-\mu_2|/\sigma) \right)
\\ &=
\frac{1}{\sigma} \sum_{i=1}^n (|x_i-\mu_2| - |x_i - \mu_1|)  \leq \frac{n}{\sigma} |\mu_1-\mu_2| = n \frac{\delta}{\sigma}.
\end{align*}
Thus, if $n \leq M \sigma/\delta$ then the likelihood ratio is at most $M$.  
If the likelihood ratio $M \leq \ln((1-p)/p)$ then with at most probability $1-p$ that $X$ is from $\LAP(\mu_1,\sigma)$. Hence, by contrapositive, only if $n > \ln(\frac{1-p}{p}) \sigma/\delta$ can we say with more than $1-p$ confidence that $X$ is from $\LAP(\mu_1,\sigma)$.
\end{proof}

For the Laplace distribution, we do not need to take expectation with respect to true distribution. We can rather get an upper bound approximation for the logarithm of the likelihood ratio $M$ by the triangle inequality shown in the proof, which cannot be achieved by other distributions, such as Gaussian. 

\begin{corollary} \label{gaussian_colloary}
Consider if a set $X$ of observations is from either $\NORMAL(\mu_1, \sigma_1)$ or $\NORMAL(\mu_2, \sigma_2)$. 
If $X$ is truly from $\NORMAL(\mu_1, \sigma_1)$, let $\eta_1 = \frac{\sigma_2}{\sigma_1}$ and $\eta_2 = \frac{\mu_2 - \mu_1}{\sigma_2},$ then in expectation at least $n > \frac{\ln(\frac{1-p}{p})}{\frac{\eta_{2}^2 - 1}{2} + \ln(\eta_{1}) + \frac{1}{2 \eta_1^2}}$ iid observations are needed for
than $(1-p) \cdot 100\%$ confidence.  
\end{corollary}

\begin{proof}
According to Case 2 in \ref{lemma-struct} we know that we need $n >  \frac{\ln \frac{1-p}{p}}{\KL(f(x \mid \mu_1) \mid \mid f(x \mid \mu_2))}$ iid observations, and 
\begin{align*}
    \KL(\NORMAL(\mu_1, \sigma_1) & \mid \mid \NORMAL(\mu_2, \sigma_2)) 
    \\ = & 
    \frac{(\mu_1 - \mu_2)^2 + \sigma_1^2 - \sigma_2^2}{2 \sigma_2^2} + \ln(\frac{\sigma_2}{\sigma_1}), 
    \\ = & 
    \frac{\eta_2^2 - 1}{2} + \ln(\eta_1) + \frac{1}{2 \eta_1^2}.
\end{align*}
Hence at least $n > \frac{\ln(\frac{1-p}{p})}{\frac{\eta_{2}^2 - 1}{2} + \ln(\eta_{1}) + \frac{1}{2 \eta_1^2}}$ iid observations are needed to conclude with more than $(1-p) \cdot 100\%$ confidence.
\end{proof}

A special case for Corollary \ref{gaussian_colloary} is when $\sigma_1 = \sigma_2 = \sigma$ and we denote $\eta = \frac{1}{\eta_2} =  \frac{\sigma}{\mu_2 - \mu_1}.$  Thus, by some simple algebra we know that if $X$ is from $\NORMAL(\mu_1, \sigma)$, then in expectation at least $n > 2\ln(\frac{1 - p}{p})\eta^2$ iid observations are needed for more than $(1-p) \cdot 100\%$ confidence. 

\section{The Basic Shift Models}
\label{sec:basic}

We first consider a case with a single sender $x \in \mathbb{R}$ and a single adversary $a \in \mathbb{R}$ in the one-dimensional domain along the line of sight between them.  
The sender will be communicating (e.g., via wifi) and may encrypt signal content, but can not hide the signal strength, and the adversary can measure this amplitude.  

The unimpeded signal presents a larger amplitude the closer it is to the sender, and so the amplitude measured by an adversary $a$ is a function of the initial magnitude $\alpha$ used by the sender, and the distance between them $|x - a|$.   
We restrict that the sender can produce signals in range $[0,M]$, with maximum signal strength $M$.  While our protocols can be adapted to any known (or learned) decay function, for simplicity in this model we will assume a linear decay of the observed signal
\[
s_{x,\alpha}(a, b=0) = \max\{0, \alpha - c \|x-a\|\}.  
\]
Here $c$ is the fixed linear decay rate, and $b = 0$ (\textbf{no person}) indicates no interference by a person.  
We next explore a few models and associated protocols so if $b=1$ (\textbf{exists person}), which can prevent an adversary from knowing the bit.  
Here we assume the sender knows this bit $b$.

\subsection{Constant Offset Interference}

In our first model, we assume that the presence of a person ($b=1$) creates a constant $\delta$ decrease in the observed signal.  
We can write the observed signal strength as
\[
s_{x,\alpha}(a, p=1) = \max\{0, \alpha - c \|x-a\| - \delta\}.  
\]


  
\begin{figure}
\includegraphics[width=\linewidth]{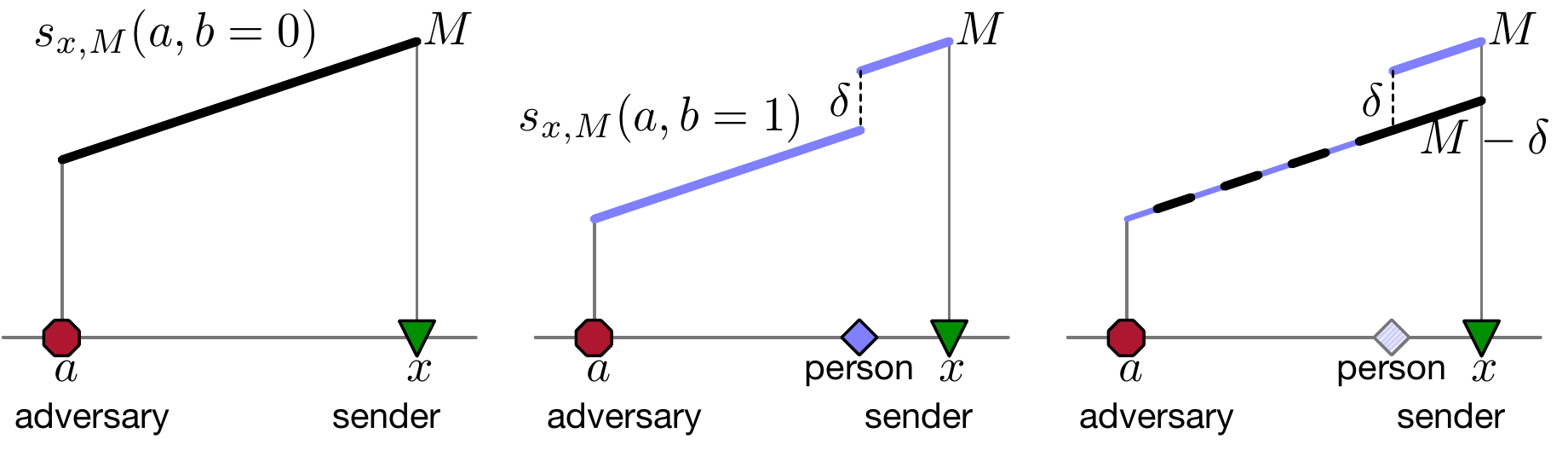}
\caption{Example of signal strength for constant offset model, and shift protocol.}
\label{fig:basic-shift}
\end{figure}

Our goal is to modulate the signal strength patter so that adversary cannot infer the person bit $b$. The \emph{Shift Protocol} is
\begin{itemize}
    \item If there is no person interfering with the signal ($b=0$), the sender emits a signal with strength $M - \delta$.
    \item If there is a person interfering with the signal ($b=1$), the sender emits a signal with strength $M$.
\end{itemize}







\begin{theorem} \label{theorem-1}
Under the Shift Protocol the adversary receives the same signal strength if a person interfering with the signal ($b=1$) or not ($b=0$), so it achieves perfect hiding. 
\end{theorem}

\begin{proof}
If $b=0$, signal strength is $M - \delta$; the signal received
\[
s = s_{x,M - \delta}(a, b = 0) = \max\{0, (M - \delta) - c \|x - a\|\}. 
\]
If $b=1$ the sent signal strength is $M$; the signal received is
\[
s' = s_{x,M}(a, b = 1) = \max\{0, M - c \|x_i - a\| -\delta \}.  
\]
And thus the observed signals $s' = s$ are identical.  
\end{proof}

\subsection{The Random Shift Protocol}
\label{sec:random-shift}


We generalize this model and protocol so the person's interference effect is not fixed, but is a random function $f$.  We consider for $f$ any \emph{known} discrete or one-sided truncated version (like truncated normal) or only with non-negative domain (like beta) continuous distributions from location-scale family\footnote{Location scale family distribution includes almost all common seen distributions like any distribution in the exponential family or some distribution not in the exponential family, say Cauchy distribution.} with a \emph{known} mean denoted as $\mu$ and variance $\sigma$, 
\begin{align*}
   \delta \sim f \;\;\;\; \text{or} \;\;\; p(\delta = y) = f_{\mu, \sigma}(y),
\end{align*}
where $p$ denote the pdf or pmf of distribution $f$.   In this model we only consider $f$ with non-negative domain.  
The \emph{Random Shift Protocol} is
%
\begin{itemize}
    \item If there is a person interfering ($b=1$), the sender emits a signal with strength $M$.
    \item If there is no person interfering ($b=0$), the sender emits a signal with strength $M-y$, where $y$ is random as $y \sim f$.  
\end{itemize}

\begin{figure}
\includegraphics[width=\linewidth]{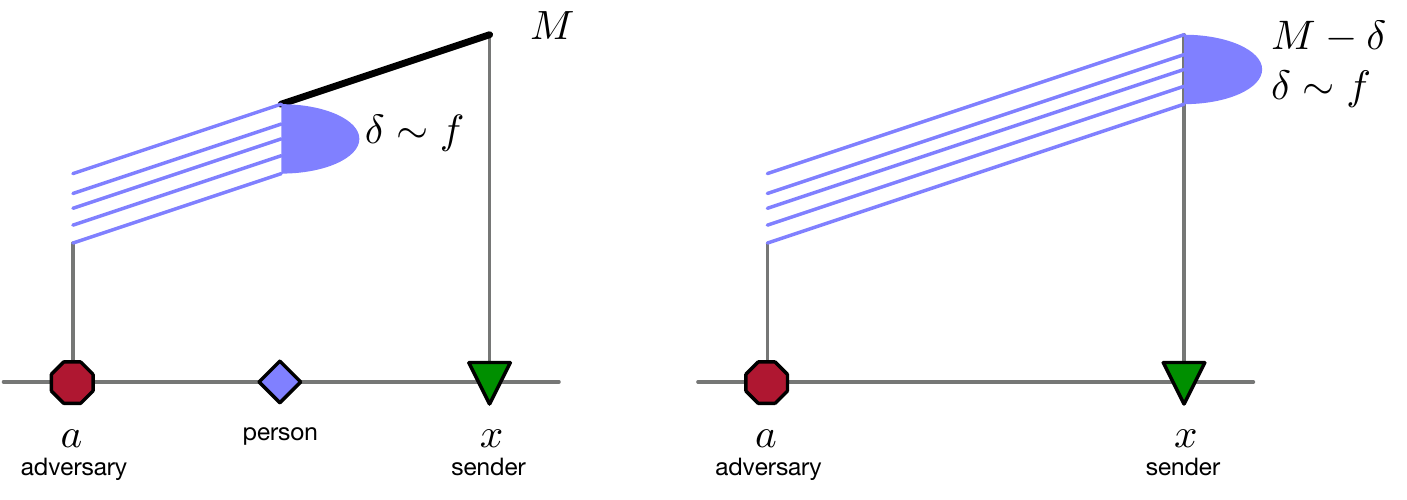}
\caption{The effect of generalized random shift, and the Generalized Random Shift Protocol.}
\label{ref:random-shift}
\end{figure}

\begin{theorem} \label{thm:rand-shift-gen}
Under the Random Shift Protocol, $S(a) \overset{d}{=} S'(a)$, so this achieves perfect hiding.  
\end{theorem}

\begin{proof}
Notice that, each $s_t(a) \in S(a)$ follows $s_t(a) = M-\delta$ where, $\delta \sim f$.  Hence for the two cases ($p=0$ and $p=1$, respectively), for each time $t$, satisfy 
\begin{align*}
     s_t(a) &= M - \delta \text{ with } \delta \sim f 
     \\
     s'_t(a) &= M - y \text{ with } y \sim f.  
\end{align*}
So we have $S(a) \overset{d}{=} S'(a)$, no matter distribution $f$.
\end{proof}



\section{Random Noise Models for Unknown Interference}
\label{sec:unknown-person}

The Basic Shift Models allowed for Case 1 (perfect hiding) in Lemma \ref{lemma-struct}.  This notably require that the sender \textbf{knows} if a person is interfering.  In this section we focus on when the sender does not know the $b$ bit, indicating interference.  In this setting, we are not able to achieve perfect hiding, and instead settle for noisy hiding (Case 2 in Lemma \ref{lemma-struct}).  

For simplicity, we analyze the constant offset interference model with reduction of $\delta$, with a linear decay with constant $c$.  Although, this can again be generalized to other decay and interference models.  


The approach will be to inject noise into signal strength chosen by the sender.  We will investigate a few types of noise, and how much is required to achieve various guarantees.  


\subsection{Laplace Noise Model}

In this setting, the sender uses signal strength chosen under a Laplacian noise model $\LAP(\mu,\sigma)$.  
If a person is interfering with the signal ($b=1$), the observed signal is reduced by $\delta$, and is equivalent to using Laplacian noise with $\LAP(\mu'_1,\sigma)$ with $\mu'_1 = \mu -\delta$.  
Then an adversary would observe data from a distribution with a reduced mean, that is from $\LAP(\mu_1,\sigma)$ or $\LAP(\mu_0,\sigma)$ with $\mu_1 = \mu'_1 - c \|x-a\|$ and $\mu_0 = \mu - c \|x-a\|$.  Ultimately, the key parameter is $\delta = \mu_0 - \mu_1 = \mu - \mu'_1$.  
A unit-less parameter $\eta = \sigma/\delta$ captures the needed characteristic of this problem; refer to this as the $\eta$-Laplace setting.


\begin{theorem}
In the $\eta$-Laplace setting, if $b=1$, 
then the adversary needs at least $n = \ln(\frac{1-p}{p}) \eta$
readings for $(1-p)\cdot100\%$ confidence in the value of $b$.  
\end{theorem}

\begin{proof}
If $b=1$, the signal strength readings received by the adversary follow $\LAP(\mu_1, \sigma)$, or $\LAP(\mu_0, \sigma)$ if $b=0$. Hence by applying Corollary \ref{laplace-corollary}, if $b=1$, the adversary needs at least $n > \ln(\frac{1-p}{p}) \eta$ readings to conclude with more than  $(1-p)\cdot100$ percent confidence the value $b$.  
\end{proof}


This approach where we can derive a bound without any probabilistic notions other than confidence is a special result of the Laplace distribution having an upper bound of their log-likelihoods being exactly $(\mu_1 - \mu_2)/\sigma$, similar to the Laplace mechanism in differential privacy~\cite{dwork2008differential}.  For other sorts of distributions (e.g., Normal), this is not the case, and we will need to state the bounds in expectation.


\subsection{Normal Noise Model}

Now we consider when the sender injects normal $\NORMAL(\mu, \sigma)$ noise into the signal strength.  Again the interference of a person ($b=1$) is modeled as decreasing the observed signal by $\delta$.  As a result an adversary would observe one of two distributions $\NORMAL(\mu_0,\sigma)$ for $b=0$ or $\NORMAL(\mu_1,\sigma)$ for $b=1$, where $\mu_0 - \mu_1 = \delta$.  Again we analyze a unit-less parameter $\eta = \sigma/\delta$ in this so-called $\eta$-Normal setting.  


\begin{theorem} \label{Gaussian-Fixed}
In the $\eta$-Normal setting, if $b=1$, then in expectation the adversary needs at least $n = 2 \ln(\frac{1-p}{p}) \eta^2$ readings for $(1-p)\cdot100\%$ in the value of $b$.  
\end{theorem}

\begin{proof}
If $b=1$, the adversary observes data from  $\NORMAL(\mu_1, \sigma)$, otherwise ($b=0$) they observe data from $\NORMAL(\mu_0, \sigma)$. 
Via Corollary \ref{gaussian_colloary}, the $\eta$-Normal setting is exactly the special case when $\sigma_1 = \sigma_2 = \sigma$. Thus if there is an interfering person, in expectation the adversary needs at least $n > 2 \ln(\frac{1-p}{p}) \eta^2$ readings for $(1-p)\cdot100\%$ confidence the existence of the person.  
\end{proof}


\subsection{Normal Noise and Normal Interference Model}
\label{sec:norm-norm}

Now we analyze when the sender injects normal noise \emph{and} the effect of the person also follows a normal distribution.  
This specific case demonstrates the general principal of how these protocols and analysis can handle noise that materialize at several spots in the broadcast, interference, and sensing process -- surely noise is introduced elsewhere as well.  

Specifically, the sender chooses signal strength from $\NORMAL(\mu, \sigma)$, and a person's interference results in a decrease in signal strength by $\NORMAL(\mu_I, \sigma_I)$.  Define $\sigma_1 = \sqrt{\sigma^2 + \sigma_I^2}$ and two unitless parameters $\eta = \sigma_1/\mu_I$ and $\eta' = \sigma_1/\sigma$.  We refer to this as the $(\eta, \eta')$-Normal setting.



\begin{theorem}\label{Gaussian-random}
In the $(\eta,\eta')$-Normal setting, if $b=1$, then in expectation the adversary needs at least  $n > \frac{\ln(\frac{1-p}{p})}{\frac{(1/\eta)^2 - 1}{2} + \log(\eta') + \frac{1}{2(\eta')^2} }$ readings to conclude with $(1-p)\cdot100\%$ the value of $b$.  
\end{theorem}

\begin{proof}
If $b=1$, then the observed signal is of the form $\NORMAL(\mu_1, \sigma_1)$ where $\mu_1 = \mu - \mu_I - c \|x-a\|$, because a convolution of two normals results in another normal with parameters $\mu-\mu_I$ and standard deviation $\sigma_1$.  Otherwise (the case $b=0$), the observed signal by the adversary would be from $\NORMAL(\mu_0,\sigma)$ with $\mu_0 = \mu - c \|x-a\|$.  


By applying Corollary \ref{gaussian_colloary}, with these normals, we determine that 
in expectation the adversary needs at least $n > \frac{\ln(\frac{1-p}{p})}{\frac{(1/\eta)^2 - 1}{2} + \log(\eta') + \frac{1}{2(\eta')^2} }$ readings to conclude with more than $(1-p)\cdot100\%$ confidence the existence of the person.
\end{proof}


\subsection{Truncated Distributions}
This above analysis formalizes how the smaller the KL divergence between two observed signals' distributions, the harder for an adversary to detect the difference.  Intuitively, when the variance of the signal sent out by the communicator is much larger than the variance of the interference, it is hard for the adversary to detect a person. 
This seems to indicate, we can set $\mu=M$ (the maximum signal strength) and set the variance ($\sigma^2$) very large, and avoid losing communication power and make it difficult to detect the presence of a person, since $\sigma \gg \delta$.  
However, the signal strength may be limited to $[0,M]$, and as a result, truncated distributions must be used.  We can set the mean $0 < \mu < M$, or more likely $2\sigma < \mu < M-2\sigma$.  
Under this restriction, the analysis in Lemma \ref{lemma-struct} can be adjusted using appropriately truncated distributions in place of a full Gaussian as in Corollary \ref{gaussian_colloary}.



\section{Models in $2$ Dimensions}
\label{sec:2d-person}

The situation in 2 dimensions with 1 person and 1 adversary can be reduced to the 1-dimensional setting.  
If the location of the person and adversary is known ($b$ is known), this maps to the Basic Shift models in Section \ref{sec:basic}.  
If the location of the person or adversary is not known ($b$ is not known), this maps to the Random Noise models in Section \ref{sec:random-shift}.

When there are 2 adversaries and 1 person, the setting is more challenging as one may be interfered with and the other not.  Assuming the adversaries collude, this allows them to immediately detect the interference, if the sent signal strength is always the same in all directions.  
This immediate detection holds even for the Random Noise model as long as the effect of the noise is the same for both adversaries.  

To circumvent this obstacle to hiding, we consider using broadcast equipment that can control directional signal strengths (as is commonplace in cell phone towers, and emerging in wifi routers).  We consider two models for directional signal strength where hiding results can exist in this setting:  very narrow band, and gradual decay.

\textbf{Very narrow band.}
In the very narrow band model, the angle in which signal is emitted from the sender is defined by an direction $\theta$, and only is detectable within a very narrow set of angles $(\theta-\tau,\theta+\tau)$, for some parameter $\tau$.  If the angle between the two adversaries from the perspective of the sender is greater than $2 \tau$, then hiding protocols exist.  
At each time point, the sender chooses a random direction to send its signal.  Then this can be observed by at most one adversary, and it again reduces to the 1 adversary case.

\textbf{Gradual decay.}
A different model does not enforce a very narrow band, but instead assumes that for a fixed direction $\theta$, the signal strength decays symmetrically in both directions as the angle of the observer becomes further from $\theta$.  Then again if the angle between the adversaries is large enough, a protocol can be designed to hide the interference of a person.  In particular, this requires the sender to know if an adversary is interfered with, and that the difference in signal strength received by the two adversaries (because of the angular decay) to be larger than the effect of the interference of the person.  

The protocol then is as follows.  
If there is no interference, direct the signal so its highest signal direction $\theta$ is directly between the two adversaries; use less than the maximum possible strength.  
If a person interferes with one adversary, then direct the signal closer towards that interfered with adversary and increase the signal strength.  If these two parameters (direction and signal strength) are chosen correctly, they can ensure both adversaries receive the same signal as without interference.

Note that in both cases, the most challenging case, which neither can overcome, is when the two adversaries are in a very similar direction from the sender, but are separated enough so a person can interfere with one but not the other.  
From one perspective, this provides a perhaps surprisingly useful setup for a pair of colluding adversaries.  
From another perspective, this model may be problematic since a person's interference with an adversary may not be so binary, and may be diffuse within a range.  If this contributes to a random distribution, some version of the noisy hiding analysis may be applicable.

Handling more than 2 adversaries is challenging in the gradual decay model because trying to solve for a setting that retains the same signal strength for 3 adversaries when there are only 2 degrees of freedom (signal strength and angular direction) is, in general, not possible.

\section{Experiments on Noisy Hiding}
\label{sec:exp}
\begin{figure}%
    \includegraphics[width=0.48\textwidth]{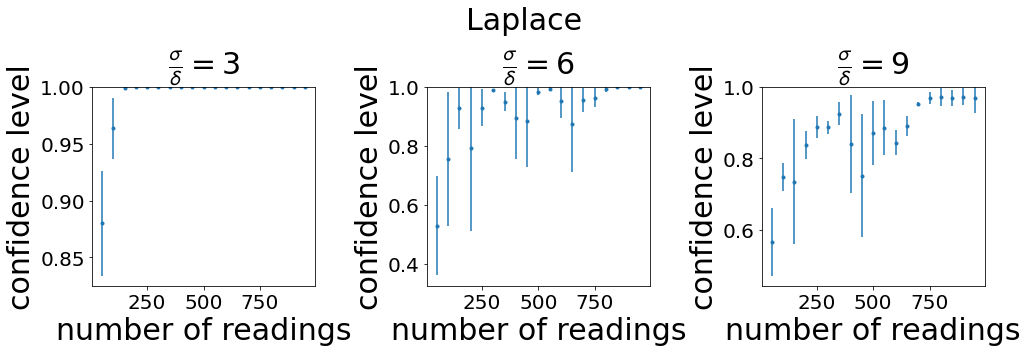} 
    \hfill
    \includegraphics[width=0.48\textwidth]{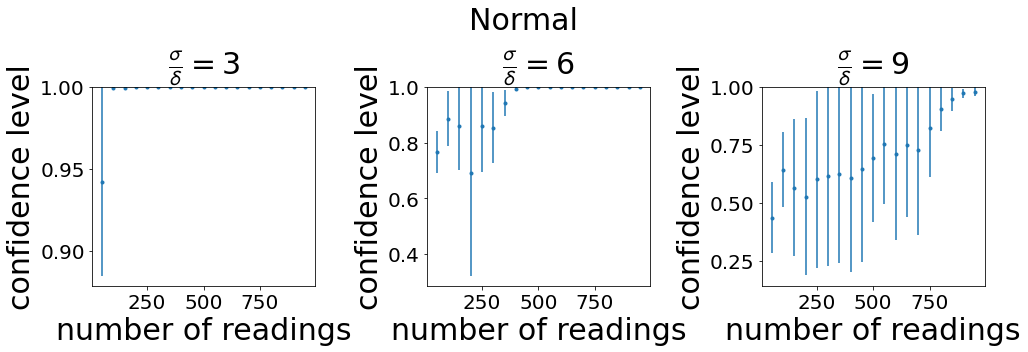}
    \caption{Statistical confidence achieved as a function of the number of readings for the Laplace and Normal noise models.  We vary the unit-less parameter $\eta = \sigma/\delta$ as $3$, $6$, or $9$.}
    \label{fig:exp}%
\end{figure}

We perform simulations on the noisy hiding model under Laplace distribution and normal distribution for different values of the ratio between standard deviation $\sigma$ and the strength of the interference effect $\delta$. 
For each value of the ratio ($\frac{\sigma}{\delta} = 3, 6, 9$) we draw $1000$ random numbers coming from the specified distribution ($\LAP(\mu, \sigma)$ or $\NORMAL(\mu,\sigma)$). 
Then we calculate the likelihoods of the random sample drawn from the true distribution $\LAP(\mu, \sigma)$ and another hypothetical distribution $\LAP(\mu - \delta, \sigma)$, and then the logarithm of the likelihood ratio between these two likelihoods. 
Our theoretical results indicates that the confidence to determine if the samples come from the true distribution rather than hypothetical distribution would be increased as the sample size increases. We perform $10$ trials for each value of the ratio, and compute the resulting confidence (via confidence = $1/(\exp(\textsf{LLR}) -1)$); and report its average and standard deviation. 
For each ratio $\frac{\sigma}{\delta}$, we vary the number of readings from $100$ to $1000$, and plot the mean and error bars (showing one standard deviation) in Figure \ref{fig:exp}.  For instance when $\sigma/\delta = 6$, then it usually requires between $500$ and $800$ readings to reach $95\%$ confidence.  
Although the Laplace distribution has a better theoretical bound (because of convenient mathematical properties), the normal empirically requires fewer trials to reach high confidence.  

\section{Discussion \& General Strategies}
\label{sec:general}
We propose new models and protocols to prevent an adversary snooping on the strength of a signal and detecting the presence of a person.  Depending on if the interference is known, the presence can either be perfectly hidden, or hidden for a number of rounds with a statistical confidence bound.  If the potentially interfering person collaborates with sender, and knows the protocol, it could potentially determine if it interferes with high confidence before a non-collaborating adversary, and shift to a perfect hiding strategy.  

The models proposed in the paper are quite general. We omitted some potential specifications -- they become tedious quickly -- but results will not deviate too much from the general theorems/properties we produce in Section \ref{sec:structural}. For instance, in observing real signal strength decay, they do not decay at a linear rate, the effect of a person blocking does decrease it but the effect is not binary (block or not block), and the fixed structures (like walls) in the environment play a role.  When these can be modeled (or learned), very similar statements can be made as simple corollaries of our main results. When they cannot be modeled effectively or precisely, the fix is essentially to add more noise that the sender is not aware of, and the extensions appear much like Theorem IV.3. 

For instance, some spatial (2-dimensional) settings with several colluding adversaries may seem hopeless, resulting in immediate detection.  But this assumes perfect information and noiseless sensing among adversaries.  Noise in the process may lead to something like the noisy hiding scenario in practice. 

These methods reduce communication rates, since a lower than maximum signal strength is used. In the noisy hiding, we bound the probability an adversary can determine the bit $b$ indicating if a person is interfering with the signal or not. Hence we must set up the mean of the distribution of the signal strength, say $\mu$, which is lower than the maximal signal strength, say $M$, and the power utilization is roughly $\mu/M$.  As discussed, how large $\mu$ can be set depending on $\sigma$ (i.e. variance of the noise), since we desire that $M-\mu > 2\sigma$ or $> 3 \sigma$ so that the truncation of a normal (or Laplace) distribution insignificantly changes the KL divergence.  In turn, this value of $\sigma$ should be roughly proportional to $\delta$ (i.e. person effect).  Since again, it is common that the interference effect of a person $\delta$ is much less than $M$, this change in power utilization should generally be tolerable.

\bibliographystyle{abbrv}
\bibliography{isit}

\begin{thebibliography}{10}

\bibitem{deep_learning}
M.~{Abbas}, M.~{Elhamshary}, H.~{Rizk}, M.~{Torki}, and M.~{Youssef}.
\newblock Wideep: Wifi-based accurate and robust indoor localization system
  using deep learning.
\newblock In {\em 2019 IEEE International Conference on Pervasive Computing and
  Communications (PerCom}, pages 1--10, 2019.

\bibitem{abrar2020quantifying}
A.~S. Abrar, N.~Patwari, and S.~K. Kasera.
\newblock Quantifying interference-assisted signal strength surveillance of
  sound vibrations, 2020.

\bibitem{AKKM}
F.~Adib, Z.~Kabelac, D.~Katabi, and R.~C. Miller.
\newblock 3d tracking via body radio reflections.
\newblock In {\em 11th {USENIX} Symposium on Networked Systems Design and
  Implementation ({NSDI} 14)}, pages 317--329, Seattle, WA, Apr. 2014. {USENIX}
  Association.

\bibitem{FD}
F.~Adib and D.~Katabi.
\newblock See through walls with wifi!
\newblock {\em SIGCOMM Comput. Commun. Rev.}, 43(4):75–86, Aug. 2013.

\bibitem{random_forest}
P.~{Bahl} and V.~N. {Padmanabhan}.
\newblock Radar: an in-building rf-based user location and tracking system.
\newblock In {\em Proceedings IEEE INFOCOM 2000. Conference on Computer
  Communications. Nineteenth Annual Joint Conference of the IEEE Computer and
  Communications Societies (Cat. No.00CH37064)}, volume~2, pages 775--784
  vol.2, 2000.

\bibitem{BBDRKB}
A.~{Baset}, C.~{Becker}, K.~{Derr}, S.~{Ramirez}, S.~{Kasera}, and
  A.~{Bhaskara}.
\newblock Towards wireless environment cognizance through incremental learning.
\newblock In {\em 2019 IEEE 16th International Conference on Mobile Ad Hoc and
  Sensor Systems (MASS)}, pages 256--264, 2019.

\bibitem{Bash_et_al}
B.~A. Bash, D.~Goeckel, D.~Towsley, and S.~Guha.
\newblock Hiding information in noise: fundamental limits of covert wireless
  communication.
\newblock {\em IEEE Communications Magazine}, 53(12):26--31, 2015.

\bibitem{Bloch_et_al}
M.~Bloch, J.~Barros, M.~R.~D. Rodrigues, and S.~W. McLaughlin.
\newblock Wireless information-theoretic security.
\newblock {\em IEEE Transactions on Information Theory}, 54(6):2515--2534,
  2008.

\bibitem{game_theory_2}
M.~Brown, B.~An, C.~Kiekintveld, F.~Ordóñez, and M.~Tambe.
\newblock An extended study on multi-objective security games.
\newblock {\em Autonomous Agents and Multi-Agent Systems}, 28, 01 2014.

\bibitem{EIT}
T.~M. Cover and J.~A. Thomas.
\newblock {\em Elements of Information Theory (Wiley Series in
  Telecommunications and Signal Processing)}.
\newblock Wiley-Interscience, USA, 2006.

\bibitem{dwork2008differential}
C.~Dwork.
\newblock Differential privacy: A survey of results.
\newblock In {\em International conference on theory and applications of models
  of computation}, pages 1--19. Springer, 2008.

\bibitem{SRD}
S.~Hussain, R.~Peters, and D.~Silver.
\newblock Using received signal strength variation for surveillance in
  residential areas.
\newblock {\em Proceedings of SPIE - The International Society for Optical
  Engineering}, 6973, 03 2008.

\bibitem{LOCAL_FREE}
O.~{Kaltiokallio}, M.~{Bocca}, and N.~{Patwari}.
\newblock Follow @grandma: Long-term device-free localization for residential
  monitoring.
\newblock In {\em 37th Annual IEEE Conference on Local Computer Networks -
  Workshops}, pages 991--998, 2012.

\bibitem{FSSP}
S.~M. Kay.
\newblock {\em Fundamentals of Statistical Signal Processing: Estimation
  Theory}.
\newblock Prentice-Hall, Inc., USA, 1993.

\bibitem{CRLBPEQS}
A.~{Moschitta} and P.~{Carbone}.
\newblock Cramer-rao lower bound for parametric estimation of quantized
  sinewaves.
\newblock In {\em Proceedings of the 21st IEEE Instrumentation and Measurement
  Technology Conference (IEEE Cat. No.04CH37510)}, volume~3, pages 1724--1729
  Vol.3, 2004.

\bibitem{game_theory_1}
P.~Paruchuri, J.~P. Pearce, M.~Tambe, F.~Ordonez, and S.~Kraus.
\newblock An efficient heuristic approach for security against multiple
  adversaries.
\newblock In {\em Proceedings of the 6th International Joint Conference on
  Autonomous Agents and Multiagent Systems}, AAMAS ’07, New York, NY, USA,
  2007. Association for Computing Machinery.

\bibitem{RFsensor}
N.~{Patwari} and J.~{Wilson}.
\newblock Rf sensor networks for device-free localization: Measurements,
  models, and algorithms.
\newblock {\em Proceedings of the IEEE}, 98(11):1961--1973, 2010.

\bibitem{monitor_breathe}
N.~{Patwari}, J.~{Wilson}, S.~{Ananthanarayanan}, S.~K. {Kasera}, and D.~R.
  {Westenskow}.
\newblock Monitoring breathing via signal strength in wireless networks.
\newblock {\em IEEE Transactions on Mobile Computing}, 13(8):1774--1786, 2014.

\bibitem{RFBO}
D.~{Rife} and R.~{Boorstyn}.
\newblock Single tone parameter estimation from discrete-time observations.
\newblock {\em IEEE Transactions on Information Theory}, 20(5):591--598, 1974.

\bibitem{TOV}
C.~Truong, L.~Oudre, and N.~Vayatis.
\newblock A review of change point detection methods.
\newblock {\em CoRR}, abs/1801.00718, 2018.

\bibitem{game_theory_3}
P.~M. Wijewardena, A.~Bhaskara, S.~K. Kasera, S.~A. Mahmud, and N.~Patwari.
\newblock A plug-n-play game theoretic framework for defending against radio
  window attacks.
\newblock WiSec '20, page 284–294, New York, NY, USA, 2020. Association for
  Computing Machinery.

\bibitem{moving_average}
M.~Youssef, M.~Mah, and A.~Agrawala.
\newblock Challenges: Device-free passive localization for wireless
  environments.
\newblock In {\em Proceedings of the 13th Annual ACM International Conference
  on Mobile Computing and Networking}, MobiCom ’07, page 222–229, New York,
  NY, USA, 2007. Association for Computing Machinery.

\bibitem{zhu2018wireless}
Y.~Zhu, Y.~Ju, B.~Wang, J.~Cryan, B.~Y. Zhao, and H.~Zheng.
\newblock Wireless side-lobe eavesdropping attacks, 2018.

\end{thebibliography}

\end{document}